\documentstyle[pre,aps]{revtex} 
\begin{document}
\draft
\title{Relaxation to steady states and dynamical exponents in deposition
models}
\author{F. D. A. Aar\~ao Reis}
\address{
Instituto de F\'\i sica, Universidade Federal Fluminense,\\
Avenida Litor\^anea s/n, 24210-340 Niter\'oi RJ, Brazil}
\date{\today}
\maketitle
\begin{abstract}
Considering some deposition models with limited mobility, we show that the
typical decay of the interface width to its saturation value is exponential,
which defines the crossover or saturation time $\tau$. We
present a method to calculate a characteristic time $\tau_0$
proportional to $\tau$ and estimate the dynamical exponent $z$.
In one dimensional substrates of lengths $L\leq 2048$, the method is applied to
the Family model, the restricted solid-on-solid ($RSOS$) model and the
ballistic deposition. Effective exponents $z_L$ converge to asymptotic values
consistent with the corresponding continuum theories. For the two-dimensional
Family model, the expected dynamic scaling hypothesis suggests a particular
definition of $\tau_0$ that leads to $z=2$, improving previous calculations
based on data collapse methods. For the two-dimensional $RSOS$ model, we
obtain $z\approx 1.6$ and $\alpha <0.4$, in agreement with recent large scale
simulations.
\end{abstract}

\pacs{PACS numbers: 05.40.+j, 05.50.+q\\
Keywords: Growth models; thin films; Universality classes; Dynamical exponent;
Relaxation}


\section{Introduction}

Statistical deposition models have attracted much attention in the last years
because they may describe real systems' features by representing the basic
growth mechanisms as simple stochastic processes, thus neglecting the details
of the microscopic interactions~\cite{barabasi}.
Some examples are the Family
model~\cite{family}, the restricted
solid-on-solid ($RSOS$) model of Kim and Kosterlitz~\cite{kk} and the
ballistic deposition ($BD$) model~\cite{vold,famvic}. The former one is
representative of the Edwards-Wilkinson ($EW$) universality class~\cite{ew} of
linear growth, while the other ones are representative of the
Kardar-Parisi-Zhang ($KPZ$) class~\cite{kpz} of non-linear growth. More complex
models involve the competition among deposition, diffusion and
aggregation~\cite{bartelt,cv,cv1} and may represent quantitatively real
systems' features.

Often one is interested in surface properties, then the main
quantity to be measured is the interface width of the deposit. If deposition
occurs in a $d$-dimensional substrate of length $L$, then the interface width
at time $t$ is defined as
\begin{equation}
W(L,t) = {\left< { {\left[ {1\over{L^d}} \sum_i{ {\left( h_i -
\overline{h}\right) }^2 } \right] }^{1/2} } \right> } 
\label{eq:1}
\end{equation}
or as
\begin{equation}
\xi(L,t) = {\left[ { \left< { { {1\over{L^d}} \sum_i{ {\left( h_i -
\overline{h}\right) }^2 } } } \right> } \right] }^{1/2} .
\label{eq:2}
\end{equation}
In Eqs. (1) and (2), $h_i$ is the height of column $i$, the bar in
$\overline{h}$ denotes a spatial average and the angular brackets denote a
configurational average, i. e., an average over many realizations of the noise.
$W$ and $\xi$ have different values but the same scaling properties, then in
the following any definition or discussion concerning $W$ is
also valid for $\xi$.

At early times, in the so-called growth regime, the interface width
increases as
\begin{equation}
W \sim t^{\beta} , 
\label{eq:3}
\end{equation}
while at long times finite-size effects lead to its saturation at
\begin{equation}
W_{sat} (L)\equiv W(L,t\to\infty ) \sim L^{\alpha} .
\label{eq:4}
\end{equation}
Relations (3) and (4) are included in the dynamic scaling relation
proposed by Family and Vicsek~\cite{famvic},
\begin{equation}
W = L^{\alpha} f\left( tL^{-z}\right) ,
\label{eq:5}
\end{equation}
in which $f$ is a scaling function and the dynamical exponent $z$ is given by
\begin{equation}
z={\alpha\over\beta} . 
\label{eq:6}
\end{equation}
In Eq. (5), the argument of the scaling function is a ratio of the deposition
time and the crossover (or saturation) time $\tau \sim L^{z}$. 

Most numerical works on deposition models focus on the calculation of
exponents $\alpha$ (Eq. 4) and $\beta$ (Eq. 3) - see e. g. the summary of
numerical results in Ch. 8 of Ref. \protect\cite{barabasi}. The calculation of
$\alpha$ requires the identification of the steady-state regime, in which $W$
saturates. Eventually, the extrapolation of effective exponents is essential,
such as in the $BD$ model~\cite{bal}. The calculation of $\beta$ depends
on the definition of the growth regime, which must be a region of
high linear correlation in a $\log{W}\times\log{t}$ plot. Finding such region
may be very difficult for $d\leq 2$ due to small lattice lengths~\cite{bal}.
The usual methods to estimate exponent $z$ are based on data collapse of the
scaled quantities appearing in Eq. (5) or on the scaling of the structure
factor $S\left( k,t\right) \equiv \langle h\left( k,t\right)
h\left( -k,t\right)\rangle$ ($h\left( k,t\right)$ is the Fourier
transformation of the surface height $h\left(
x,t\right)$)~\cite{plischke1,plischke2,alanissila}. In a recent work, Kim et al
presented a method to estimate $z$ from the exponential decay of a relaxation
function, starting the growth process from a sinusoidal
surface~\cite{kim}. In $d=1$, they obtained estimates for the Family, the
$RSOS$ and the Krug models which were consistent with the expected universality
classes~\cite{kim}. However, for the most common problem of deposition in a
flat surface, a precise definition of the crossover time $\tau$ and a method to
estimate $z$ accurately are still lacking.

It is also important to recall that there are controversies on the
universality classes of various statistical growth models even in
$d=1$~\cite{evans}. At this point, the development of reliable techniques to
estimate critical exponents is also helpful.

The aim of this work is to analyze the convergence of the scaling function in
Eq. (5) to an asymptotic constant value (consistently with Eq. 4) and to
propose and test methods to estimate crossover times and the exponent $z$.

First we will consider the Family, the
$RSOS$ and the $BD$ models in $d=1$. We will show that the typical decay of
$W$ to its saturation value is exponential, which uniquely defines the
crossover or saturation time $\tau$. Then we will present a method to
calculate a characteristic time $\tau_0$ that is proportional to $\tau$, and
we will estimate the exponent $z$ from $\tau_0$. The reliability of the method
will be proved with comparisons with exact results for the corresponding
hydrodynamic equations.

We will also consider the Family and the $RSOS$ models in $d=2$.
Results for the $RSOS$ model confirm the exponential decay of $W$ to
the saturation value and give $z\approx 1.6$. Estimates of the exponent
$\alpha$, obtained from $W_{sat}$ data, are in good agreement with the
prediction of Marinari et al ~\cite{marinari} that $\alpha<0.4$ for $KPZ$ in
$d=2$. For the Family model, we will consider the expected dynamic scaling
hypothesis, which involves a $W\sim\log{L}$ growth, and we will show
results consistent with $z=2$. A recent numerical work~\cite{landau} showed
large discrepancies from that value, using data collapse techniques, which
proves the relevance of the methods presented here.

The rest of this paper is organized as follows. In Sec II we present the
results of simulations in $d=1$ for the Family, the $RSOS$ and the $BD$
models, show the exponential decay of $W$ to its saturation value and the
methods to calculate $z$. In Sec. III we present the simulations' results of
the Family and the $RSOS$ models in $d=2$, including estimates of exponent
$\alpha$ for the $RSOS$ model. In Sec. IV we summarize our results and present
our conclusions.

\section{Definition of the models and results in one-dimensional substrates}

First we recall the definition of the models studied here.

In the Family model~\cite{family}, a column of the deposit is
randomly chosen and the incident particle searches for the smallest height in
its neighborhood. If no neighboring column has a smaller height than the column
of incidence, the particle sticks at the top of this one. Otherwise, it sticks
at the top of the column with the smallest height among the neighbors. If two
or more neighbors have the same minimum height, the sticking position is
randomly chosen among them.

In the $RSOS$ model~\cite{kk}, the incident particle may stick at the top
of the column of incidence if the differences of heights of all pairs of
neighboring columns do not exceed ${\Delta H}_{MAX} = 1$. Otherwise, the
aggregation attempt is rejected.

In $BD$, the incident particle follows a straight trajectory perpendicular to
the surface and sticks upon first contact with a nearest neighbor occupied
site. It leads to the formation of a porous deposit, contrary to the previous
solid-on-solid models.

In all models, a time unit corresponds to the deposition of $L$ particles.
We simulated them in one-dimensional substrates with lengths
$L=2^n$, with integer $7\leq n\leq 11$ ($128\leq L\leq 2048$). For each $L$,
typically ${10}^4$ different deposits were generated, and very long
steady-state regions were observed.

In order to obtain reliable estimates of $W_{sat}$, we computed it over
different time ranges to ensure that it was fluctuating around an average
value, and not increasing systematically in time (the latter possibility would
suggest that the steady-state region was not attained yet). $W_{sat}$ was
obtained with accuracy smaller than $0.2\%$ in all cases, which enables the
calculation of the deviation from $W_{sat}$,
\begin{equation}
{\Delta W}\left( t\right) \equiv W_{sat} - W\left( t\right) ,
\label{eq:7}
\end{equation}
also with good accuracy.

In Fig. 1a we show $\log{\Delta W}\times t$ for the Family model in
$L=1024$, and in Fig. 1b we show $\log{\Delta W}\times t$ for the $RSOS$ and
$BD$ models, also in $L=1024$. In Figs. 2a and 2b we show $\log{\Delta
\xi}\times t$ in the same systems. It clearly shows an exponential decay of
$W$ and $\xi$ to $W_{sat}$ and $\xi_{sat}$, in the form
\begin{equation}
{\Delta W}\left( t\right) = A\exp{\left( -t/\tau \right)} ,
\label{eq:8}
\end{equation}
with constants $A$ and $\tau$.
The estimates of relaxation times shown
in Figs. 1 and 2 were obtained from fits of the data in regions of high linear
correlation. For the other lengths $L$, similar linear behaviors in
$\log{\Delta W}\times t$ plots were obtained.

The calculation of $\tau$ from those plots has some disadvantages. For
instance, $\tau$ has to be estimated in different time ranges, in order to
avoid spurious effects of an arbitrary choice of the fitting region. Thus we
searched for an alternative method, as described below, which is not based on
similar data fits.

The scaling relation (5) and Eq. (8) imply
that the amplitude $A$ of Eq. (8) is proportional to $L^{\alpha}$.
Consequently, when approaching the steady-state region, the scaling function
of Eq. (5) may be written as
\begin{equation}
f(x) = a - b\exp{\left( -x\right)} ,
\label{eq:9}
\end{equation}
with constant $a$ and $b$. If a characteristic time $\tau_0$ is defined from
\begin{equation}
{W\left( \tau_0 \right) } = k W_{sat} ,
\label{eq:10}
\end{equation}
with constant $k$, then
\begin{equation}
\tau_0 = \tau \ln{\left[ {b\over {a\left( 1-k\right) }} \right] } ,
\label{eq:11}
\end{equation}
i. e., $\tau_0$ is proportional to the crossover time $\tau$.
The estimates of $\tau_0$ from Eq. (10) always had higher accuracy
than the relaxation times obtained from $\log{\Delta W}\times
t$ plots (typically for a factor $3$).

We considered two possible values of the constant $k$ in Eq. (10), $k_1 =
1-1/e = 0.6321\dots$ and $k_2 = 0.8$. The first
choice would give $\tau_0 = \tau$ if $a=b$ (Eqs. 9 and 11), which seems to be
approximately valid for $BD$. The effective exponents $z_L$ were obtained from
\begin{equation}
z_L = { \ln \left[ \tau_0\left( L\right) /\tau_0\left( L/2\right)\right]
\over \ln 2 } ,
\label{eq:12}
\end{equation}
and must be independent of the particular choice of the constant $k$.

In Figs. 3a and 3b we show $z_L$ versus $1/L$ for the three models using
$k=k1$ and $k=k_2$, respectively, in the calculation of $\tau_0$ (Eq. 10). $W$
data were used, but the effective exponents obtained from $\xi$ are nearly the
same. The uncertainties in $W_{sat}$ are responsible for the error bars, thus
results with $k=k_1$ (larger $\Delta W$) are more accurate.

In all cases shown in Figs. 3a and 3b, $z_L$ converges to
the exactly known values as $L\to\infty$. In the Family and in the $RSOS$
models, $z=2$~\cite{ew} and $z=3/2$~\cite{kpz} are recovered with small
finite-size corrections. In $BD$, $1/L$ corrections to $z_L$ lead to expected
$z=3/2$~\cite{kpz} asymptotic value. It is interesting to notice that these
corrections in $BD$ are weaker than the corrections found in effective
exponents $\beta_L$ and $\alpha_L$ in Ref. \protect\cite{bal}, which proves the
relevance of developing methods to estimate $z$ independently. Moreover, this
result complements that systematic study of $BD$ in $d=1$, which
discussed the discrepancies in numerical estimates of critical exponents and
the controversies on the equivalence of $BD$ and the $KPZ$ theory~\cite{bal}.

\section{Results in two-dimensional substrates}

First we consider the $RSOS$ model in $d=2$. Simulations were performed for
lengths $32\leq L\leq 512$, with a small number of realizations (nearly
${10}^3$) in the largest lattices. The exponential decay of  ${\Delta W}\left(
t\right)$ is confirmed in Fig. 4a for lattices of length $L=512$. In Fig. 4b
we show $z_L$ versus $1/L$, which gives the asymptotic estimate $z=1.60\pm
0.05$. This result is consistent with several previous estimates of critical
exponents for that model~\cite{barabasi}.

Using $W_{sat}$ data, we were also able to obtain finite-size estimates of the
exponent $\alpha$:
\begin{equation}
\alpha_L = { \ln \left[W_\infty\left( L\right) /W_\infty\left( L/2\right)\right]
\over \ln 2 } .
\label{eq:13}
\end{equation}
In Figs. 5a and 5b we show $\alpha_L$ versus $1/L$ obtained from $W$ and
$\xi$ data, respectively. Both sets of data strongly suggest an
asymptotic exponent $\alpha <0.4$, which agrees with simulations of Marinari et
al~\cite{marinari}, who obtained $\alpha = 0.393\pm 0.003$. However, it
disagrees with the proposal of $\alpha = 0.4$ based on an operator
product expansion~\cite{lassig}, which suggests further investigations on
these lines.

For the Family model, it is expected the dynamic scaling
relation~\cite{forrest}
\begin{equation}
W^2 = A\ln{\left[ L{f\left( t/\tau \right)} \right] }
\label{eq:14}
\end{equation}
where $f$ is a scaling function and $A$ is constant. As $x\to\infty$, it is
expected that $f(x)\to a$, with constant $a$.
The characteristic time $\tau_0$ defined from Eq. (10) is not proportional to
$\tau$ anymore. Then, in order to define a characteristic time $\tau_1$ which
is proportional to $\tau$ in this system, we consider
\begin{equation}
{\delta W^2} \equiv {W_{sat}}^2 - {W\left( \tau_1 \right) }^2 = C ,
\label{eq:15}
\end{equation}
with constant $C$. 

The characteristic times $\tau_1$ were calculated using Eq. (15) with
$C=0.03$. The effective exponents $z_L$
were also obtained from Eq. (12). In Figs. 6a and 6b we show $z_L$ versus $1/L$
obtained from $W$ and $\xi$ data, respectively. In both cases, $z_L$
converges to $z=2$ as $L\to\infty$, in agreement with the $EW$ theory.
Our result is significantly better than the value $z\approx 1.63$, obtained in
Ref. \protect\cite{landau} using data collapse methods, which again proves the
reliability of our method.

\section{Summary and conclusion}

We proposed and tested methods to estimate crossover times of interface
width scaling and the dynamic exponent $z$.
We showed that the typical decays of $W$ or $\xi$
to their saturation values are exponential, which uniquely defines the
crossover or saturation times $\tau$. In order to estimate $z$ with accuracy,
we used a characteristic time $\tau_0$ that is proportional to $\tau$,
according to the Family-Viczek scaling. The reliability of the
method was proved by testing it with the Family, the $RSOS$ and $BD$ models
in $1+1$ dimensions, in which dynamic exponents of the $EW$ and the $KPZ$
theory were obtained. The Family and the $RSOS$ models in $d=2$ were also
considered. For the $RSOS$ model, the exponential decay of $W$ is confirmed,
and $z\approx 1.6$ was obtained. Moreover, estimates of saturation widths
suggest that $\alpha <0.4$, in agreement with a recent
prediction of large scale simulations~\cite{marinari}. For the
two-dimensional Family model, the expected dynamic scaling
hypothesis lead to an alternative definition of the characteristic time
$\tau_0$, and the exact value $z=2$ was recovered.

The relevance of our methods to estimate dynamical exponents is also proved by
comparisons with previous results for related systems. For instance, using 
data collapse of relaxation functions and the expected scaling of $W$ at
$t=L$, $z=1.55$ was obtained for Eden clusters~\cite{plischke1} and $z=1.57$
for the single-step model~\cite{plischke2} in $d=1$. Both systems are in the
$KPZ$ class (exact $z=1.5$). These difficulties were pointed out in the paper
of Kim et al~\cite{kim}, who proposed a method to measure $z$ from the scaling
of a relaxation function and an initial sinusoidal surface. Although the
efficiency of their method was clearly stated, working with a flat surface
is the most usual initial situation in theoretical works and is more suitable
for real applications. Moreover, with our method, one is able to calculate
three exponents, $\alpha$, $z$ and $\beta$, using the same set of data. The
relevance of each one to determine the universality class of a particular
model may depend on several factors, such as the strength of finite-size
corrections. Thus we expect that this work may be useful for further studies
of surface growth models.

\acknowledgements

This work was partially supported by CNPq and FINEP (brazilian agencies).
The author thanks the Department of Theoretical Physics at Oxford University,
where part of this work was done, for the hospitality.

\begin{figure}
\caption{Time evolution of the deviation from the saturation width for: (a) the
$1d$ Family model; (b) the $1d$ $RSOS$ and $BD$ models (slower decay for
$RSOS$). The linear fit of the data in $2000<t<6000$ is shown for the Family
model. Relaxation times calculated from such fits are shown for the three
models.}
\label{1}
\end{figure}

\begin{figure}
\caption{Time evolution of the deviation from the saturation width for the
same models of Fig. 1, with $\xi$ instead of $W$.}
\label{2}
\end{figure}

\begin{figure}
\caption{Effective exponents $z_L$ versus $1/L$ in $d=1$, obtained using
(a) $k=k_1=1-1/e$ and (b) $k=k_2=0.8$, for the Family model (squares), the
$RSOS$ model (triangles) and the $BD$ model (crosses). Data for the $RSOS$
model are slightly shifted to the left.}
\label{3}
\end{figure}

\begin{figure}
\caption{(a) Time evolution of the deviation from the saturation width for the
$RSOS$ model in $d=2$; (b) effective exponents $z_L$ versus $1/L$ for the same
model, using $W$ data and $k=k_1$.}
\label{4}
\end{figure}

\begin{figure}
\caption{Effective exponents $\alpha_L$ versus $1/L$ for the $RSOS$ model in
$d=2$, obtained from (a) $W$ and (b) $\xi$ data.}
\label{5}
\end{figure}

\begin{figure}
\caption{Effective exponents $z_L$ versus $1/L$ for the $EW$ model in $d=2$,
obtained from (a) $W$ and (b) $\xi$ data.
}
\label{6}
\end{figure}

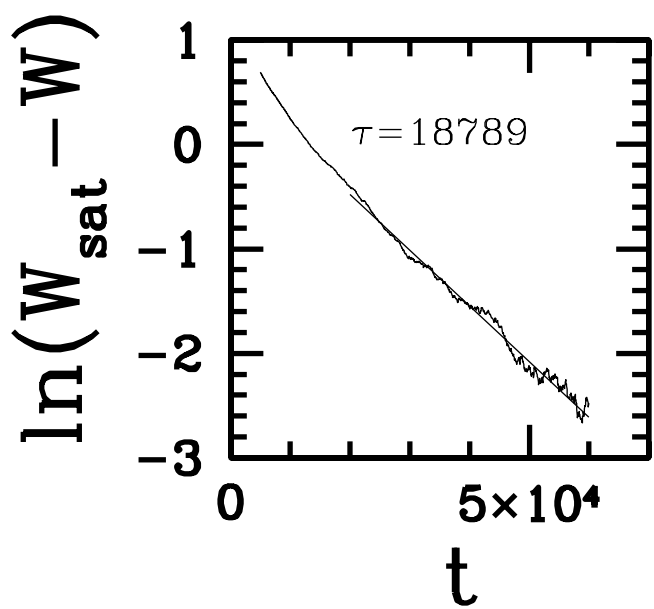
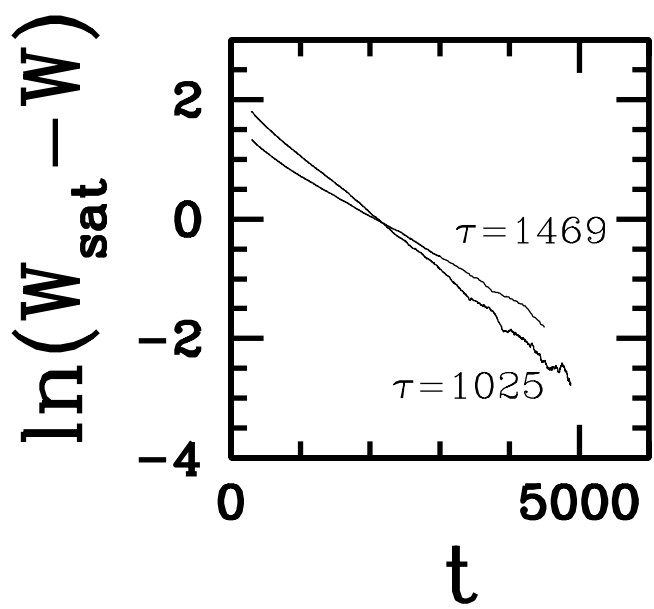
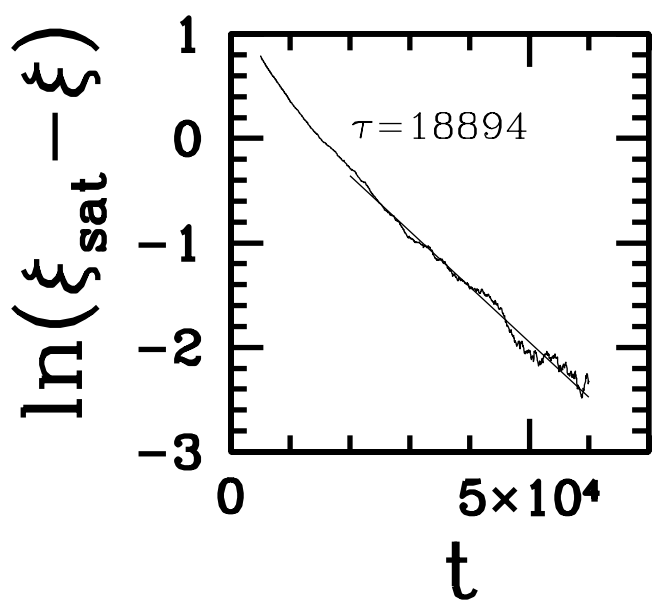
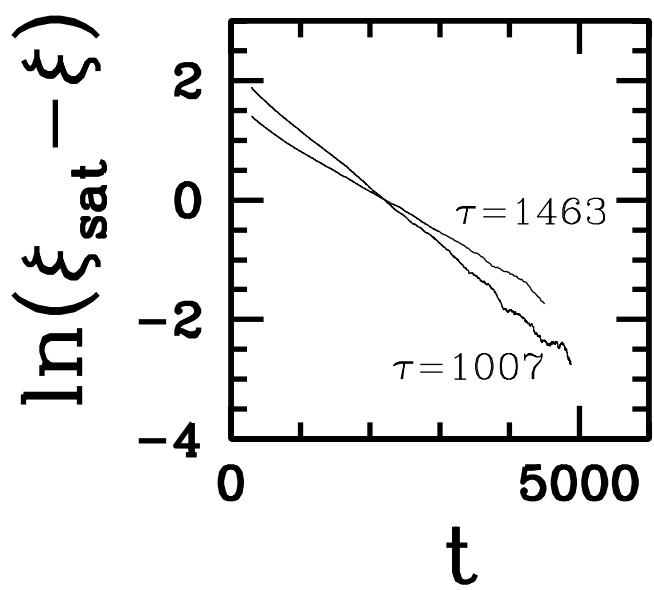
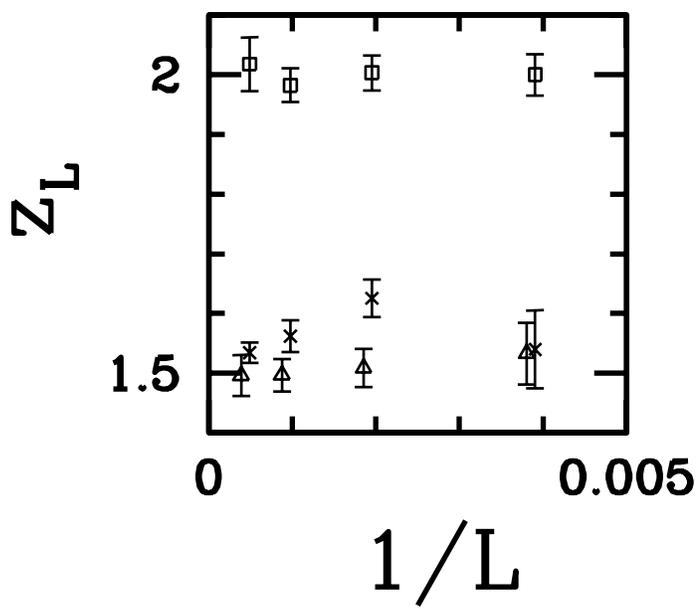
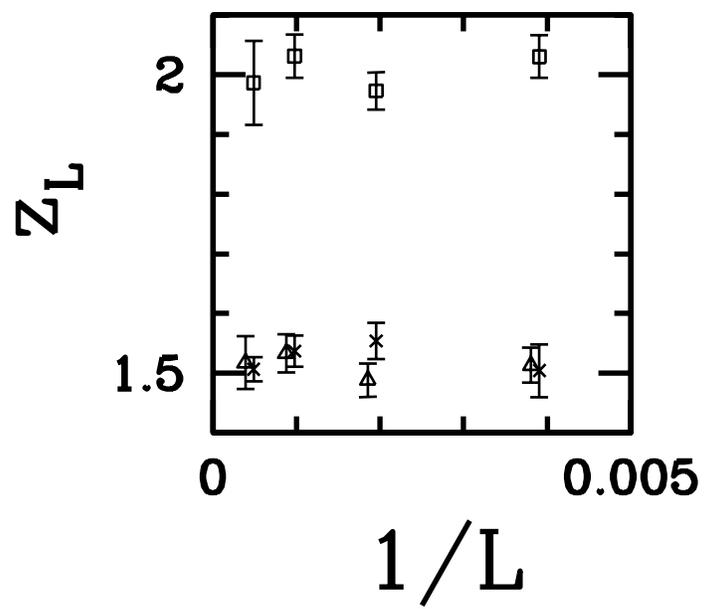
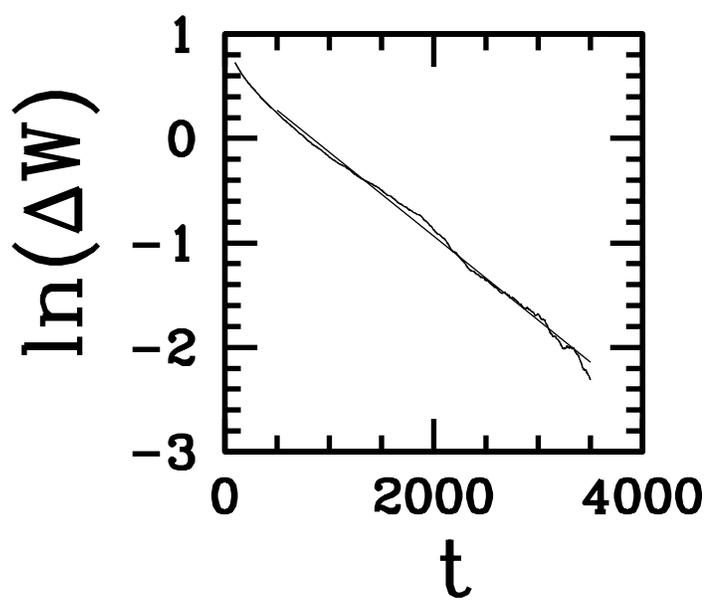
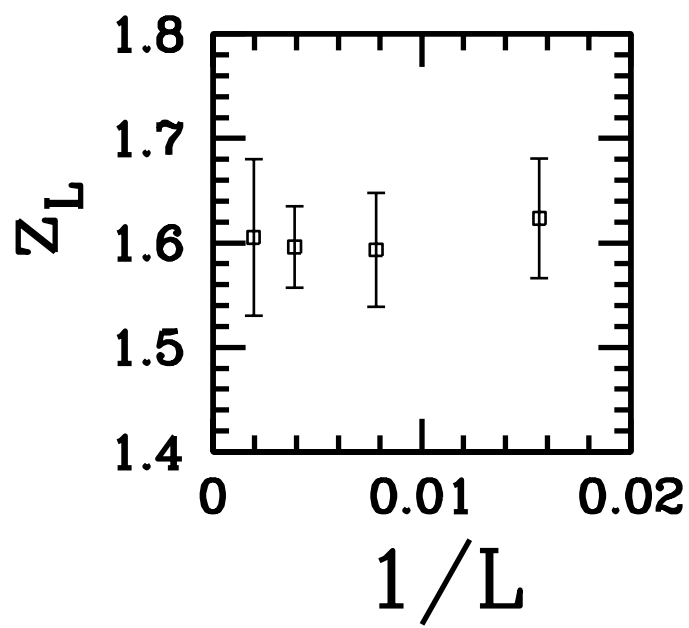
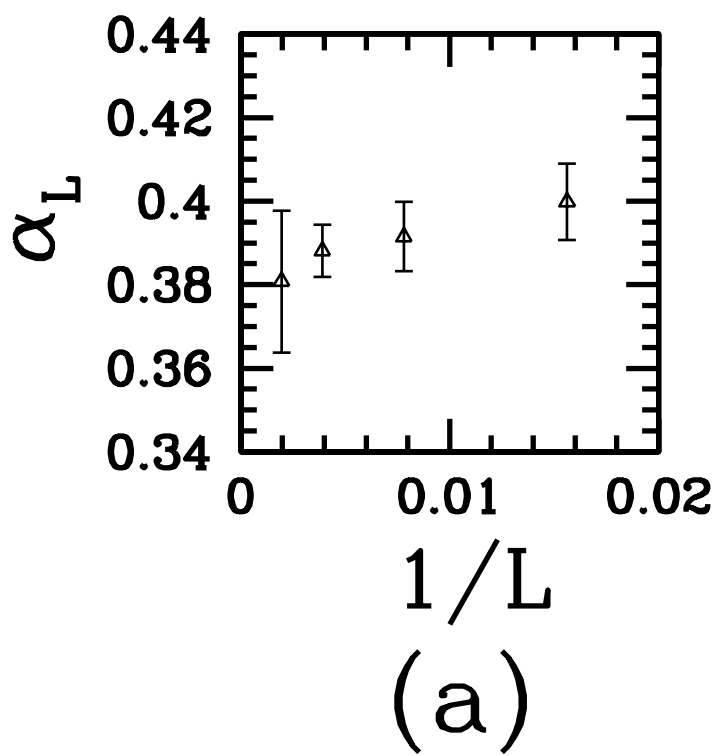
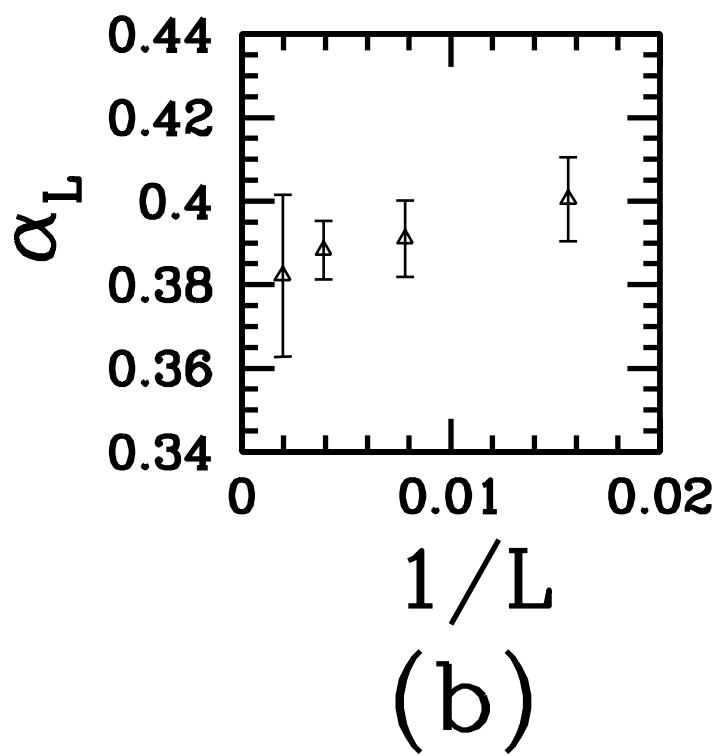
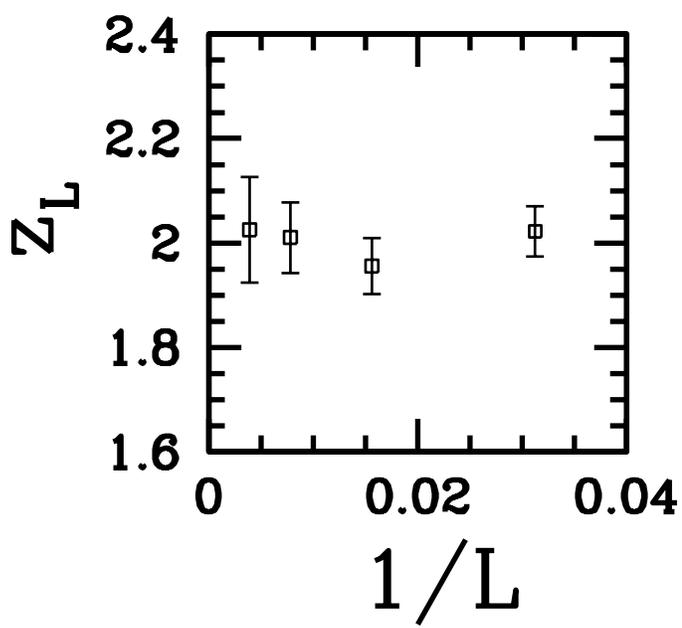
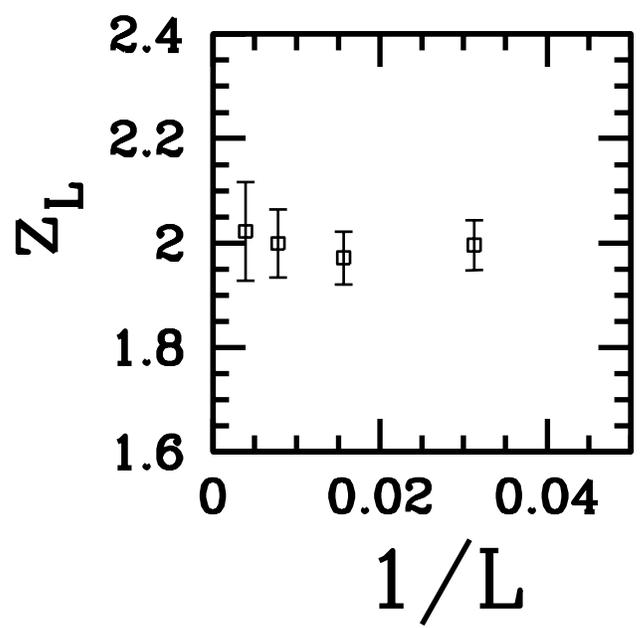
\end{document}